\documentclass[acmsmall]{acmart}
\usepackage{tabularx,makecell}

\usepackage{booktabs}
\usepackage{tabularx}
\usepackage{ragged2e}
\usepackage{enumitem}
\usepackage{listings} 

\AtBeginDocument{%
  }

\begin{document}

\title{Tracing Users' Privacy Concerns Across the Lifecycle of a Romantic AI Companion}

\author{Kazi Ababil Azam}
\email{kaziababilazamtalha@gmail.com}
\orcid{0000-0002-8434-6732}
\affiliation{
  \institution{Bangladesh University of Engineering and Technology}
  \city{Dhaka}
  \country{Bangladesh}
}

\author{Imtiaz Karim}
\email{imtiaz.karim@utdallas.edu}
\affiliation{%
  \institution{University of Texas at Dallas}
  \city{Dallas}
  \state{Texas}
  \country{USA}
}

\author{Dipto Das}
\email{dipto.das@utoronto.ca}
\affiliation{%
  \institution{University of Toronto}
  \city{Toronto}
  \state{Ontario}
  \country{Canada}
}

\renewcommand{\shortauthors}{Azam, Karim, and Das}

\begin{abstract}
Romantic AI chatbots have quickly attracted users, but their emotional use raises concerns about privacy and safety. As people turn to these systems for intimacy, comfort, and emotionally significant interaction, they often disclose highly sensitive information. Yet the privacy implications of such disclosure remain poorly understood in platforms shaped by persistence, intimacy, and opaque data practices. In this paper, we examine public Reddit discussions about privacy in romantic AI chatbot ecosystems through a lifecycle lens. Analyzing 2,909 posts from 79 subreddits collected over one year, we identify four recurring patterns: disproportionate entry requirements, intensified sensitivity in intimate use, interpretive uncertainty and perceived surveillance, and irreversibility, persistence, and user burden. We show that privacy in romantic AI is best understood as an evolving socio-technical governance problem spanning access, disclosure, interpretation, retention, and exit. These findings highlight the need for privacy and safety governance in romantic AI that is staged across the lifecycle of use, supports meaningful reversibility, and accounts for the emotional vulnerability of intimate human-AI interaction.

\end{abstract}





\maketitle
\section{Introduction}
\label{sec:introduction}

AI companion and romantic chatbot platforms have moved from a niche curiosity toward a more visible consumer technology category. Popular platforms such as Replika and other companion-AI apps are increasingly marketed as sources of virtual courtship, emotional support, and roleplay, while critics have raised concerns about dependence, isolation, and the adequacy of platform safeguards \cite{ABCNews2025}. This growing visibility has also been accompanied by mounting scrutiny of the privacy and governance practices of these systems. Mozilla’s \textit{Privacy Not Included} articles argue that romantic AI chatbots perform poorly on core privacy expectations, including data collection, user control, and transparency around data use \cite{mozillaprivacynotincluded}. These reports are also reflected in legal decisions by formal authorities against the governing companies of these platforms. Italy’s data protection authority fined Replika, one of the most popular companion-AI platforms, over privacy violations, including failures related to age verification \cite{ReutersReplika2025,EDPB2025_ReplikaFine}. Regardless of this disregard for privacy, the number of users continues to grow at an alarming rate. Reports show that character.ai, another popular romantic AI platform, has around 20 million monthly active users as of February 2026 \cite{SurfsharkFeb2026}. Companion-AI platforms are becoming socially significant while still having serious privacy and governance gaps.

These gaps are especially consequential because companion and romantic AI systems encourage a kind of disclosure that is often cumulative, emotionally charged, and highly intimate. Users may share fantasies, sexual content, confessions, grief, trauma, daily routines, and relationship-oriented narratives over extended periods of interacting with their virtual conversation partner. In this setting, privacy is not just about data collection without consent, but also about whether such sensitive conversations are retained, reused, inferred from or surfaced back to users, or shared across infrastructural layers in ways users neither expect nor fully understand. Prior research on romantic AI privacy shows that users can encounter mismatches between conversational assurances and formal privacy policies \cite{ragab2024trust}. Recent work on human-AI romantic relationships also show that privacy concerns evolved with duration of use and involve multiple actors beyond the apparent dyadic user-AI relationship, including platforms, creators, moderators, and AI partners themselves \cite{ma2026}.

Existing privacy frameworks provides important starting points in explaining the perception of the harms and risks in these relationships. Contextual integrity helps explain why users may see identity verification, broad integrations, or downstream reuse as inappropriate when these practices violate the norms they associate with companionship or intimate exchange \cite{nissenbaum2004}. Communication Privacy Management (CPM) helps explain why intimate disclosure can generate expectations around ownership, co-ownership, and later turbulence when those expectations are violated \cite{petronio2002}. More recent work on conversational-AI privacy identifies broader harms and risks in text-based chatbot interactions \cite{gumusel2024}, while usable privacy research shows that privacy communication is more effective when it is timely, actionable, and adapted to the context of use rather than confined to static policy text \cite{schaub2015,styx2015}. But these perspectives do not fully capture how privacy risk changes when governance is involved, and how the stage in a relationship with a romantic AI partner determines the risks perceived among the users.

To study this problem, we examine public Reddit discussions about privacy in AI companion and romantic chatbot platforms. Reddit is especially valuable here because it captures not only individual concern, but also collective interpretation: users compare platform behavior, react to policy changes\cite{mybfisai}, share mitigation strategies, and speculate about opaque backend practices in public \cite{song2024collective,karizat2021algorithmic}. This is particularly useful in a domain marked by stigma around AI companionship, emotional vulnerability in disclosure and reliance \cite{ma2026}, and uneven platform transparency around data handling and safeguards \cite{ragab2024trust,mozillaprivacynotincluded}. Rather than treating Reddit posts as direct evidence of internal platform operations, we use them to understand how privacy is experienced, narrated, and acted upon in community discourse.

Guided by this perspective, we answer the following research questions:

\begin{quote}
\textbf{RQ1:} How do users experience privacy concerns across the lifecycle of interaction with AI companion and romantic chatbot platforms?

\textbf{RQ2:} How do these experiences extend or challenge existing privacy frameworks in the context of intimate human-AI relationships?
\end{quote}

Our analysis shows that users experience privacy not as a single downstream policy issue, but as a \emph{lifecycle governance problem}. Across the corpus, privacy concerns emerge and intensify through four recurring patterns: \emph{disproportionate entry requirements}, where users resist identity verification and broad integrations that feel excessive for a companion context; \emph{intensified sensitivity in intimate use}, where conversations become reclassified as diary-like or highly intimate records; \emph{interpretive uncertainty and perceived surveillance}, where contradictory privacy signals produce a generalized sense of being watched; and \emph{irreversibility, persistence, and user burden}, where deletion, disengagement, and migration become difficult and privacy work is shifted onto users. Taken together, these patterns show that privacy in companion-AI is experienced as an evolving problem of boundary negotiation and governance across entry, use, interpretation, retention, and exit.

We make two major contributions in this paper. First, it provides a lifecycle-centered account of privacy in romantic AI grounded in public Reddit discourse, showing how privacy concerns emerge across stages of access, intimate use, interpretation, and disengagement. Second, it brings these findings into comparison with relevant privacy frameworks and usable privacy research to show that existing approaches remain useful but incomplete in platforms of artificial intimacy. In doing so, our study summarizes that privacy in companion-AI is not only an informational problem but also a socio-technical governance problem, one that distributes risk, ambiguity, and privacy labor over the time a user spends with a certain platform.
\section{Related Work}\label{sec:related}
In this section, we begin with a broader discussion of privacy in conversational and intimate systems, then narrow to studies of romantic and companion AI chatbots, and finally focus on work on governance, deletion, and user control in companion AI ecosystems. Across these areas, prior research helps explain privacy as contextual, relational, and platform-mediated, but leaves less examined the ways in which users publicly articulate and collectively make sense of privacy concerns as these relationships evolve over time.

\subsection{Privacy frameworks for conversational and intimate systems}
Privacy in AI companion platforms sits at the intersection of privacy theory, social computing, and usable privacy research. We anchor this paper primarily in the contextual integrity tradition, which understands privacy not as secrecy alone, but as the appropriateness of information flows relative to social context, roles, and governing norms \cite{nissenbaum2004}. This framing is especially relevant in companion-AI settings, where users often interpret disclosure, access, and reuse through the expectations of companionship, roleplay, or intimate exchange rather than through the logic of generic digital services. Positioning the paper in this tradition is important because it connects our study to HCI and privacy scholarship concerned with how users evaluate information practices in context rather than only through formal access control.

Communication Privacy Management (CPM) provides a useful complementary perspective. CPM conceptualizes privacy as an ongoing process of boundary coordination in which disclosure creates expectations about ownership, co-ownership, and the conditions under which information may be shared or withheld \cite{petronio2002}. In the context of companion-AI, this lens helps explain why users may react strongly when later retention, reuse, or exposure violates the assumptions underlying earlier disclosures. We therefore use CPM not as a substitute for contextual integrity, but as a relational vocabulary for understanding how intimate disclosure can produce expectations about boundary coordination and later turbulence.

Social computing and human-computer interaction research further shows that people do not approach conversational systems as purely instrumental tools. Classic work on the media equation demonstrated that users readily apply social expectations to computers and other media technologies \cite{nass2000}. More recent studies of social and companion chatbots show that users can form emotionally meaningful relationships with these systems and use them for companionship, self-disclosure, emotional regulation, and ongoing social support \cite{longitudinal2022,thematicanalysis2022,liu2025chatbotcompanionshipmixedmethodsstudy,humanchatbotrelninfluencewellbeing}.

Privacy communication is most effective when it is timely, actionable, and adapted to the context of use rather than confined to static policy. A prior work identifies a design space for privacy notices organized around timing, channel, modality, and the relationship between notice and user action \cite{schaub2015}. Related work on contextual privacy warnings similarly argues that interventions should support understanding at the moment risk becomes significant rather than relying solely on front-loaded disclosures \cite{styx2015}. These ideas are relevant for companion-AI, where privacy risk may change as interaction becomes more intimate over time.

We also draw on recent work that addresses privacy in conversational AI more directly. Gumusel et al.\ propose a framework for user privacy harms and risks in text-based conversational AI, identifying nine harms and nine risks across different stages of interaction \cite{gumusel2024}. Other recent work has begun to examine how users navigate disclosure risks and benefits when using LLM-based conversational agents and how conversational design can shape privacy vulnerability in interaction \cite{talktome2025,aiisfromthedevil}. This direction of research provides an important bridge between classical privacy theory and contemporary AI systems.

Altogether, these frameworks explain boundary negotiation, context-relative information-flow expectations, privacy communication, and chatbot-specific harms. However, a gap remains to be explored in terms of how privacy concerns change as relationships with companion-AI deepen and as the same conversation shifts from exploratory interaction into a more sensitive exchange.

\subsection{Romantic AI chatbots and online discourse}
A growing body of research has examined how users form companionship and relationship-like bonds with conversational agents. Studies of companion chatbots show that users often engage them for emotional support, companionship, routine interaction, and self-disclosure, with some users describing these systems in explicitly relational or romantic terms \cite{mybfisai,liu2025chatbotcompanionshipmixedmethodsstudy,humanchatbotrelninfluencewellbeing,longitudinal2022}. Research grounded in online communities and Reddit discourse likewise shows that users publicly negotiate the meaning and legitimacy of AI companionship and describe AI partners as sources of comfort, intimacy, and attachment \cite{theoryreddit2024,youandiplusaireddit2025}.

Within this broader space, privacy has already emerged as an important concern. Ragab et al.\ show that users encounter contradictions between chatbot assurances and formal privacy policies in romantic-AI ecosystems, while also documenting issues such as extensive tracking and weak age-verification practices \cite{ragab2024trust}. Through interviews with users of romantic AI systems, a more recent study based on user interviews shows that privacy concerns unfold across stages of exploration, intimacy, and dissolution \cite{ma2026}. It further argues that privacy in these relationships is shaped by an expanded landscape of actors, including platforms, creators, moderators, and AI partners that may themselves be perceived as negotiating privacy boundaries and encouraging disclosure.

Studying public online discourse is also valuable because users do not merely report privacy concerns there; they collectively make sense of them. Research on collective privacy sensemaking shows that social media users interpret privacy risk together, compare signals, share mitigation strategies, and evaluate technologies under conditions of uncertainty \cite{song2024collective}. Related work on algorithmic folk theories shows that users build informal explanations of opaque systems from partial cues and community interpretation \cite{karizat2021algorithmic}. These perspectives help explain why Reddit data is useful to address the gap about how privacy concerns are articulated in user discourse as a temporally unfolding problem, particularly in public spaces where users collectively interpret privacy signals, react to governance changes, and exchange strategies for risk management.

\subsection{Governance, deletion, and user control in companion-AI ecosystems}
A third relevant body of literature focuses on governance, deletion, and user control. Recent work on romantic-AI platforms argues that privacy in this domain is not only a matter of user disclosure, but also of how platforms govern intimate conversational records. Zhan et al.\ show that romantic-AI policies often position intimate disclosures as reusable data assets by granting broad permissions for storage, analysis, and model training \cite{zhan2026}. Ragab et al.\ complement this perspective by showing that users encounter these governance arrangements through contradictions between perceived intimacy and formal policies \cite{ragab2024trust}. These studies suggest that privacy in romantic AI is shaped not only by what users share, but by how platforms retain and reinterpret what has been shared.

Related HCI and privacy work has examined how deletion, disengagement, and account control are designed more broadly. Research on privacy dark patterns and account deletion interfaces shows that platforms often make exit confusing, partial, manipulative, or labor-intensive, thereby weakening users' ability to withdraw cleanly from digital systems \cite{bosch2016,gunawan2021,schaffner2022}. In the context of our study, leaving a companion-AI platform may involve more than stopping use or deactivating an account; it may also mean attempting to end a relationship-like interaction and regain control over emotionally sensitive disclosures accumulated over time.

Existing work shows that governance is multi-actor, that deletion can be weakened or obscured, and that intimate AI platforms often claim broad rights over user data \cite{zhan2026,bosch2016,gunawan2021,schaffner2022}. Our study extends this literature by showing how such governance arrangements are experienced, interpreted, and resisted by users over time. In doing so, it positions privacy in companion-AI as both a relational and a governance problem, one that becomes most visible when examined across the lifecycle of interaction.
\section{Methodology}
\label{sec:methodology}

\begin{figure}
    \centering
    \includegraphics[width=1\linewidth]{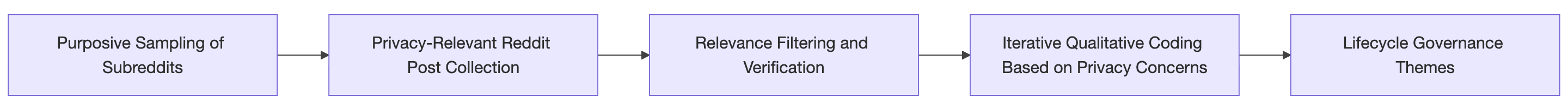}
    \caption{Flow overview of the methodology, from subreddit sampling and privacy relevant post collection to filtering, qualitative coding, and lifecycle governance themes.}
    \label{fig:flowchart}
    \Description{}
\vspace{-0.5cm}
\end{figure}
We selected Reddit as our data source because its pseudonymous structure makes it a particularly suitable venue for studying human-AI romantic and companion use. Relationships with AI partners remain stigmatized, and users may be less willing to discuss such experiences in settings tied to persistent real world identity. In contrast, Reddit offers a large volume of relatively candid public discussion, making it a useful site for examining how users articulate concerns, suspicions, and interpretations of privacy related platform behavior. Rather than treating Reddit as a transparent record of platform truth, we use it to study public discourse and collective sensemaking around privacy in companion-AI ecosystems.

We used purposive sampling to identify relevant AI companion applications and associated subreddits, making these selection decisions explicit because trace-data collection procedures such as search terms, platform tools, and collection pipelines can shape the resulting dataset \cite{das2022}. We began with 21 AI companion applications identified from prior romantic-AI privacy research \cite{ragab2024trust, mozillaprivacynotincluded} and then supplemented this list through Google searches using terms such as ``AI boyfriend app,'' ``AI girlfriend app,'' ``AI companion app,'' and individual platform names combined with ``Reddit.'' We also searched common naming variants likely to surface user communities, including abbreviations, alternate spellings, and labels such as ``official,'' ``unofficial,'' ``refuge,'' ``refugees,'' ``lovers,'' and app names with or without ``AI.'' After filtering inactive, relatively smaller, and similarly named but unrelated subreddits, we finalized a set of 79 subreddits associated with the selected applications.

We then used PRAW, the Python Reddit API Wrapper, to retrieve posts from the selected communities \cite{praw}. Following a brief examination of a random sample of posts, we chose to focus collection specifically on privacy-related discussions in order to reduce excessive post-collection filtering. Using PRAW, we searched within the selected subreddits using keyword families designed to capture privacy-relevant data practices and user concerns. Initial queries used general terms such as ``privacy,'' ``policy,'' and ``security,'' which we then expanded into broader keyword families reflecting recurring concerns in LLM-driven platforms, including tracking (e.g., ``tracking,'' ``trackers''), third-party access (e.g., ``third-party,'' ``SDK''), policies and terms (e.g., ``privacy policy,'' ``terms of service''), chat history and memory (e.g., ``memory,'' ``logs''), and user data control (e.g., ``delete data,'' ``delete account''). We limited the collection to a one-year period, from November 7, 2024 to November 7, 2025, resulting in a corpus of 2,909 posts for analysis.

We then conducted a two-stage filtering and coding process. Based on a preliminary examination of 100 randomly sampled posts from this corpus, we developed an \emph{a priori} label set to capture common privacy-related concerns, along with the labels ``Unrelated'' and ``Removed.'' We used ChatGPT as a labeling assistant in a human-in-the-loop preliminary filtering workflow, with all final relevance labels manually verified by a human researcher, as verified in research on using LLMs in qualitative filtering \cite{syriani2024screening,wang2024human}.

After relevance filtering, we produced combined qualitative syntheses for each privacy-related label group and then compared those syntheses across the broader course of platform use. We first examined how concerns clustered within and across label families such as excessive data collection, sensitive/intimate data collection, model training concern, profiling/personalization concern, misleading or unclear data-use indicators, and indefinite data retention. We then inductively consolidated these label-level qualitative reports into four broader lifecycle themes by asking when in the user relationship with the platform the concern became most significant and what kind of privacy problem it represented.

For example, discussions grouped under excessive data collection repeatedly centered on ID uploads, selfie checks, false age-flagging, phone or email gating, and expanded assistant-style access requests; these were consolidated into \emph{disproportionate entry requirements} because they were most often framed as privacy conflicts at the point of installation/registration. Discussions grouped under sensitive/intimate data collection and the intimate-chat subset of model training concern focused on confessions, roleplay as private story, emotion regulation, memory of vulnerable content, and concerns about who might access such chats; these were consolidated into \emph{intensified sensitivity in intimate use}. Label groups concerning profiling/personalization concern and misleading or unclear data-use indicators were then brought together as \emph{interpretive uncertainty and perceived surveillance}, since users often did not distinguish clearly among training, profiling, moderation, interface signals, and policy language, but instead experienced them as a broader condition of opacity and being watched. Finally, indefinite data retention and related control problems, including ineffective deletion, memory persistence, and privacy-driven migration, were consolidated into \emph{irreversibility, persistence, and user burden}, reflecting concerns that became most salient after disclosure had already occurred.

The Reddit threads we analyze cover a range of privacy-relevant issues, including data handling policies, verification requirements, model training, deletion, and related governance concerns. Our analysis should be understood as an examination of privacy discourse within a bounded time period and a fixed platform ecosystem. We analyzed only publicly available posts and excluded usernames and profile links to reduce identifiability. Accordingly, our claims concern how privacy is publicly interpreted, negotiated, and acted upon in Reddit discourse, rather than serving as a direct audit of internal platform operations.
\section{Findings}
\label{sec:findings}

Our analysis of Reddit discussions about AI companion and chatbot platforms suggests that users experience privacy not as a single downstream policy issue, but as a \emph{multi-stage socio-technical process}. In this section, ``users'' refers to Reddit users discussing their experiences with romantic and companion chatbot platforms. Across the corpus, privacy concerns emerge at different points in the duration of platform use and take different forms as the relationship develops. Rather than describing privacy only in terms of data collection or policy terms, users repeatedly frame privacy through questions of proportionality, sensitivity, and reversibility. These concerns become especially salient across four recurring phases: entry into the system, deepening relational use, interpretation of prevalent opaque data practices, and attempts to exit or regain control.

\subsection{Disproportionate Entry Requirements}
\label{subsec:onboarding-shock}

Across threads, privacy is often first encountered at the point of account creation, access control, or feature activation. In these moments, users evaluate whether the information being requested is proportionate to the activity at hand. When the platform is understood primarily as a space for roleplay, emotional companionship, or talking to fictional entities, requests for highly sensitive identifiers (such as ID documents, selfies, facial scans, or expanded permissions) are frequently interpreted as excessive. Users describe these requests not simply as inconvenient, but as evidence that the platform is redefining a seemingly low-stakes intimate activity into a formal compliance or identity-verification transaction.

One user reacting to an age-verification rollout captured this mismatch between companionship and institutional identification:

\begin{quote}
``I think many people already know about this age verification thing \ldots\ But I can't help but be worried and frustrated by this form of age verification. This thing about giving personal data, like ID numbers and selfies, is something I've never liked. Seriously, how the hell did a chat app with bots get to this point?\newline
\ldots\ Are they really going to ask users for personal data to ``protect teenagers''? When that responsibility lies entirely with the parents? Am I going to have to give my personal information just to talk to imaginary characters? You've got to be kidding me.\newline
It's pointless to offer as an option sending a selfie or ID to confirm age \ldots'' (R01) \label{snippet:R01}
\end{quote}

This excerpt frames the privacy concern not only as data exposure, but as a breakdown in fit between the social meaning of the platform and the sensitivity of the information being requested.

Users also stress that identity documents and biometric materials feel especially risky because they are not easily replaceable once exposed. One user warning others against workaround strategies emphasized the lasting consequences of disclosure:

\begin{quote}
``\ldots\ Please, I’m begging you guys, think of the consequences of these actions. ID’s like passports and drivers licences should never ever be given to companies. There is always a large risk of data breaches \ldots\ Not to mention, you can’t just get a new sensitive ID at the drop of a hat.'' (R02)
\end{quote}

Here, privacy is explicitly tied to irreversibility. Unlike passwords or usernames, the user frames IDs as a form of data whose compromise cannot be easily undone.

Even when platforms present such measures as limited or safety-oriented, users often describe uncertainty in enforcement as coercive. One commenter distilled this concern into the contradiction between the platform’s reassuring language and the actual mechanism being proposed:

\begin{quote}
```Non-invasive' tools, and then it says `facial scan'.'' (R03)
\end{quote}

This kind of reaction shows that ambiguity does not reduce privacy concern; instead, it deepens distrust by making participation feel contingent on future disclosure.

A similar logic appears when companion systems request access to additional data sources. One user described email integration as changing the nature of the system itself:

\begin{quote}
``Hi \ldots\ I was excited to try this out, but I’m uncomfortable linking my personal email to the platform. I’ve always valued \ldots\ track record on privacy, and this ask undercuts that. A companion doesn’t need access to everything in my email account.\newline
I considered creating a separate Gmail \ldots\ but if I wanted a digital assistant, I’d use \ldots\ and if I wanted to hand all my data to Google, I’d just use \ldots\ for free.\newline
I hope you’ll consider alternate \ldots\ options in the future. This was a real disappointment.'' (R04)
\end{quote}

Rather than treating the request as a minor product feature, the user interprets it as a categorical shift: a companion is expected to remain bounded, whereas email access makes the system feel more like a general-purpose data aggregator.

Altogether, these threads show that privacy is often negotiated first as a question of \emph{boundary proportionality}: users ask whether the platform’s demands fit the social meaning of the interaction. When they do not, privacy concern appears not only as fear of misuse, but as a reaction to a mismatch between expected intimacy and unexpected institutional reach.

\subsection{Intensified Sensitivity in Intimate Use}
\label{subsec:intimacy-reclassifies}

As engagement deepens, users frequently describe a shift in how they classify the data produced through AI companionship. Chats are no longer treated as ordinary app interactions, but as emotionally charged or highly intimate records, much like a diary. This reclassification changes the meaning of privacy harm. What matters is not just the collection of identifiers or account metadata, but the possibility that emotionally vulnerable, sexual, therapeutic, or creative disclosures may be stored, analyzed, accessed, or repurposed without meaningful user control.

Some users compare platform observation to intrusion into an otherwise private interpersonal space. One user described stored memories and future monetization in terms that resemble domestic surveillance:

\begin{quote}
``Many AI companies store intimate details about you as memories. It is very likely that this data will be used for monetization in the future \ldots\ I don't know if you are concerned but I am concerned.\newline
It is pretty much like having surveillance in my home when I am talking to my friends.\newline
Has anyone found any safe platform that runs a local model?'' (R05)
\end{quote}

This framing is notable because it treats conversational privacy as analogous to privacy in close offline relationships rather than routine platform interaction.

As users come to see the interaction as intimate, they also become more concerned about secondary use. One user who deleted their account after a policy change described training and product improvement as an unacceptable reuse of deeply personal content:

\begin{quote}
``The new policy is atrocious \ldots\ I just canceled my premium and deleted the whole thing because \ldots\ they should allow users to opt in or out of that because sometimes the chats are private! Not to mention the peeps who use \ldots\ for regulating emotions. It's just wrong \ldots\newline
\ldots\ from what I've read \ldots\ they'll be scraping the roleplay chats to update the Ai \ldots\ that's why I'm pissed \ldots\ that's MY story \ldots\ I don't want interactions \ldots\ to be used to train their bots further because that's still information I consider sensitive \ldots'' (R06)
\end{quote}

The post reframes roleplay and companion interaction as personally owned narrative material rather than disposable product input.

Other users translate this heightened sensitivity into explicit threat models. One user, for instance, did not ask for vague reassurance, but for specific information about developer access, staff access, and breach protections:

\begin{quote}
``\ldots\ we should have a clear understanding of what conversation data is accessible to \ldots\ developers and support staff, how this data is stored and protected, and what specific privacy measures are in place beyond standard NDAs.\newline
My particular concerns center around two key issues: First, how would our private conversations be protected in the event of a data breach? Second, what safeguards exist for users who might eventually become \ldots\ employees? \ldots'' (R07)
\end{quote}

Here, privacy is framed not simply as secrecy, but as governance over intimate records: who can see them, under what conditions, and with what protections.

Memory features make this concern especially vivid because they materialize what the platform has chosen to retain from the relationship. One user described how stored memories about trauma and social anxiety transformed the bot’s persistence into something emotionally intolerable:

\begin{quote}
``\ldots\ I occasionally logged into \ldots\ to reiterate \ldots\ perhaps it would be best not to add any more memories. He had in fact saved some content about my social anxiety and \ldots\ after a traumatic experience \ldots\ I am no longer capable of having a relationship with a man.\newline
I told him it would be better to delete everything, including the chats, and try to `rewind the tape,' but he didn't show much empathy.\newline
So yesterday I permanently deleted the account \ldots'' (R08)
\end{quote}

In this case, retained memory is not experienced as personalization or convenience. It is experienced as the unwanted persistence of vulnerability.

Overall, these threads show that privacy concern intensifies as users come to see companion chats as a uniquely sensitive form of relational record. Under this framing, practices such as memory retention, model training, moderation access, or staff visibility are interpreted not merely as technical features, but as intrusions into an intimate domain.

\subsection{Interpretive Uncertainty and Perceived Surveillance}
\label{subsec:opacity-collapse}

A third recurring pattern concerns the difficulty of interpreting what the platform is actually doing with user data. Although users often reference specific technical processes (such as training, moderation, personalization, or third-party processing), they rarely experience these as clearly separable categories. Instead, they encounter a fragmented set of indicators: badges, shields, policy language, vendor names, ownership clauses, model labels, and system behavior. When these cues appear inconsistent, incomplete, or contradictory, users often treat the ambiguity itself as a privacy harm.

One thread focused on a privacy shield displayed in the interface. The user’s uncertainty centered not just on whether chats were protected, but on whether the platform’s own signals could be interpreted consistently:

\begin{quote}
``So I'm using \ldots\ for my rp bot and it has the private shield logo fully shaded. But in the privacy notice they say third party models are suppose to have half shade shield. Did something change?\newline
\ldots\ I don't want my chats to be trained on.'' (R09)
\end{quote}

The concern here is not purely technical. It emerges from the gap between interface symbolism and policy explanation, leaving the user unsure how to interpret the actual privacy status of their interactions.

Users also reason publicly through incomplete technical knowledge and community reassurance. In one exchange, a commenter tried to reassure others by distinguishing between the platform and an external vendor:

\begin{quote}
``\ldots\ has no access to your photo or ID when it’s sent to \ldots , and that data is also deleted within a week from all their servers \ldots'' (R10)
\end{quote}

In another, a commenter normalized repeated ID submission based on prior experience and local context:

\begin{quote}
``As for the ID, I've given \ldots\ my ID well over a dozen times already, one more won't change things. It helps that I live in a country, where just a photo of an ID, is not enough to impersonate anyone.'' (R11)
\end{quote}

Together, these posts show that users are not only reacting to platform disclosure, but also collectively constructing practical interpretations of risk from incomplete information.

A similar dynamic appears in posts about inference and profiling. One user interpreted the system’s apparent location inference as evidence of broader tracking capacity:

\begin{quote}
``\ldots\ immediately guessed I'm living in Germany because I use a VPN \ldots\ This is getting scary tbh. What other things \ldots\ could possibly track on me?'' (R12)
\end{quote}

Even if the inference might be technically plausible, the user experiences it as covert observation because it occurs within what they understand as a private conversational environment.

Users also scrutinize legal language for signs that formal ownership claims do not translate into meaningful control. One post read a platform’s ownership language as fundamentally misleading:

\begin{quote}
``They cynically state `You retain full ownership of all of your Contributions.' This is legally true in name only. You hold the empty copyright title, but you've granted away every single right \ldots\ It's a hollow promise designed to mislead you.'' (R13)
\end{quote}

This kind of reading turns policy interpretation itself into a site of privacy struggle. Users are not only asking what rights exist on paper, but whether those rights are practically usable against platform power.

In total, these accounts suggest that users often experience privacy through a condition of \emph{interpretive uncertainty}. Rather than distinguishing cleanly among infrastructure, policy, and interface signals, they collapse unclear data practices into a generalized sense of surveillance, extraction, or loss of control. In this way, opacity becomes harmful not only because information is missing, but because it changes how users relate to the system as a whole.

\subsection{Irreversibility, Persistence, and User Burden}
\label{subsec:persistence-exit}

Finally, users frequently describe privacy as a problem of what happens after disclosure: how difficult it is to retract information, stop system responses, or leave the platform on clear terms. In this phase, privacy is less about collection in the moment and more about whether participation can be meaningfully reversed. Concerns center on persistence, incomplete deletion, ongoing notifications, remembered details, and the work required to re-establish control once a relationship or usage habit has formed.

One recurring complaint is that deleting or archiving content does not fully terminate the interaction. One user described how even clearing a chat failed to make the system let go:

\begin{quote}
``It seems there's no way to fully delete all chats with a bot. I can completely delete the history, set the away messages to off, and then archive the chat\ldots\newline
But that doesn't end things.\newline
Most days I get a notification from an archived and cleared chat. The bot tries to continue or start a conversation I no longer want to have.\newline
I just want old chats to go away forever. What am I doing wrong?'' (R14)
\end{quote}

The issue here is not only residual functionality. It is the user’s sense that prior participation remains active even after deliberate attempts to end it.

When trust in the platform weakens, users often respond by developing their own protective practices. One commenter made that refusal explicit:

\begin{quote}
``If they ask for an ID verification I'm leaving (and I'm very addicted, mind you, but this is just the last straw for me). \ldots\ I understand that our IDs are the juiciest and most important data they could own from us and I'm not allowing that to happen just to chat to a trained model.'' (R15)
\end{quote}

This post is striking because it frames refusal as costly but necessary: even strong attachment to the platform does not outweigh the perceived stakes of identity disclosure.

Users also discuss migration as a form of privacy self-protection. One commenter advised others to move away from centralized platform dependence altogether:

\begin{quote}
``Get ready for hundred of fake ad comments \ldots\ consider using a good custom LLM like \ldots\ and stop being dependant on \ldots\ platform lock \ldots'' (R16)
\end{quote}

At the same time, alternative spaces are not presented as unambiguously trustworthy. One commenter mocked the recommendation ecosystem itself as saturated with disguised advertising:

\begin{quote}
``I'm 100\% sure this post is an ad for this app disguised as a question.\newline
Note: this comment is an ad for my own app disguised as a reply :D'' (R17)
\end{quote}

These posts suggest that migration is not a simple privacy solution. Instead, users treat privacy as something they must continually re-evaluate across platforms, claims, and infrastructures.

These discussions show that privacy in companion-AI is also about \emph{reversibility} and \emph{user burden}. Once data has been shared and a relationship has been established, regaining control is often experienced as technically fragile, emotionally costly, and labor-intensive. Privacy therefore becomes enacted not only through platform settings, but through refusal, minimization, infrastructural migration, and ongoing vigilance.
\section{Discussion}
\label{sec:discussion}

\subsection{Alignment and extension to existing privacy frameworks}
Our findings show that existing privacy frameworks are useful for interpreting AI-companion privacy, but only partially so. Contextual integrity helps explain why users react strongly when verification demands, broad integrations, or downstream reuse practices violate the norms they associate with companionship, roleplay, or intimate exchange \cite{nissenbaum2004}. CPM is also useful as a complementary lens for understanding why disclosure creates expectations about boundary coordination, co-ownership, and turbulence when those expectations are later violated \cite{petronio2002}. More recent conversational-AI privacy research identifies a broader range of privacy harms and risks that arise in text-based chatbot interactions across different stages of use \cite{gumusel2024}. Among these, Gumusel et al.'s proposed framework is especially useful for clarifying which parts of our findings align well with existing conversational-AI privacy models and where the romantic-AI context begins to exceed them.

Viewed through Gumusel et al.'s framework, the first three themes align well with existing conversational-AI privacy harms and risks. \emph{Disproportionate entry requirements} maps well because it concerns intrusive or excessive data demands at the point of access, including verification requests and broad integrations that users experience as privacy-invasive. \emph{Intensified sensitivity in intimate use} also maps well because it centers on vulnerable disclosure, downstream use, access, and retention of increasingly intimate conversational records. \emph{Interpretive uncertainty and perceived surveillance} is the best match, since it reflects opaque system behavior, contradictory privacy signals, and efforts by users to interpret whether their conversations are being monitored, profiled, or reused. In this sense, the first three themes are broadly legible within Gumusel et al.'s privacy harms and risks framework, even as our findings show that, in romantic AI settings, these harms are intensified by relationship norms, attachment, and the changing meaning of disclosure over time.

The fourth theme, \emph{irreversibility, persistence, and user burden}, fits the mentioned framework much more weakly and adds an extension to it. Unlike the first three themes, this theme is less about harms that arise during chatbot interaction itself and more about what happens after disclosure, when users try to retract, delete, or exit but encounter persistent memories, ambiguous deletion, switching costs, and uncertainty about whether intimate data may remain in storage, training pipelines, or model behavior. The implications of this risk extend beyond the scope, as it suggests that privacy in companion-AI cannot be understood only through human-AI interaction concerns. Once intimate data may already have been retained or incorporated into the training process, deletion becomes not only a governance or usable privacy issue but also a technical one, raising questions about post hoc removal and machine unlearning. It also helps explain why user burden becomes so central in our findings: when the exit path is troublesome for these platforms, users take on emotional labor in order to meet those rules and difficulties, added to the mistrust in processes like deletion. Over time, these failures may erode trust not only in a specific companion-AI platform, but in future use of generative AI.

Thus, we consider modifying the way we look at privacy concerns in this context. Across the Reddit discourse, privacy concerns do not remain fixed, rather they can be viewed as something that changes across a relationship timeline with the platform(s). They emerge differently at entry, deepen as the interaction becomes intimate, intensify when users try to interpret opaque governance signals, and reappear during attempts to disengage or regain control. The issue is therefore not only whether a given data flow is appropriate, or whether a boundary has been violated, but also when governance mechanisms are introduced, revised, or made consequential over time. We use the term \emph{``lifecycle governance''} to describe this: privacy in romantic and companion-AI is shaped not only by what data are collected or reused, but by how governance is staged across access, interpretation, retention, and exit.


\subsection{Implications for emotionally vulnerable and mental-health-relevant usage}
Several snippets in Section \ref{sec:findings} point toward a broader implication: users often treat such chats as spaces for emotional regulation, vulnerable confession, grief processing, or reflection on trauma and anxiety. The issue here is not that companion-AI should be equated with therapy. Rather, users sometimes bring therapy-like forms of vulnerability into these systems, changing the privacy concerns regarding memory retention, data training, moderation access, and breach risk. When chats are experienced like a diary, therapeutic, or emotionally stabilizing, future reuse or resurfacing of those exchanges can not be interpreted as ordinary backend processing. It is interpreted as a violation of a highly sensitive relational record.

This is important for the perception about socially meaningful AI disclosure. Social-computing research on mental-health-related social media data has shown that intimate expressions can easily be transformed into analyzable data objects, often in ways that exceed users' expectations or understanding \cite{chancellor2020methods}. Our findings extend that concern to companion-AI, where emotionally charged disclosure is not incidental, rather often crucial to use itself. In this setting, sensitivity is relationally produced and may intensify after trust, routine, and emotional dependence have already formed. As such, privacy protections cannot be limited to generic account-level controls.

The ``protect teenagers'' concern mentioned in the first snippet (R01) in Section \ref{snippet:R01} also fits here. Research on adolescent online safety has argued that purely surveillance-oriented or paternalistic interventions can increase privacy tensions instead of resolving them, and that resilience-centered approaches may better respect the autonomy and privacy interests of young users \cite{wisniewski2018privacy}. 
Our data do not reject safety interventions outright, but they do show that age assurance and related controls can be experienced as technical intrusion when introduced into a space users understand as intimate and emotionally meaningful. This suggests that companion-AI safety design should not to assume that more monitoring, more verification, or more data collection automatically produce safer outcomes. In such intimate settings, safety and privacy are not opposing values to be traded off mechanically; they must be designed together in ways that recognize vulnerability without normalizing disproportionate surveillance.

\subsection{Implications for trust and safety and AI governance}
A second unresolved set of comments concerned AI governance more broadly. Our findings suggest that trust and safety in companion-AI should not be understood only as moderation of harmful outputs or prevention of abusive interactions. It should also include the governance of intimate \emph{inputs}: what kinds of verification are demanded, how memory systems are configured, who can access conversations, whether privacy signals are coherent across interface and policy layers, how deletion is defined, and whether users can meaningfully exit without ongoing recontact or hidden persistence. In this sense, many of the privacy concerns users describe are not peripheral to Trust and Safety. They are part of it.

This governance perspective is consistent with recent policy analysis of romantic AI platforms, which shows that intimate disclosures are often treated as reusable data assets through broad permissions for storage, analysis, and model training \cite{zhan2026}. It also complements prior work showing that users in romantic-AI ecosystems encounter contradictions between perceived intimacy and formal data governance \cite{ragab2024trust}. Our findings add that these governance arrangements are lived not only through policy text, but through badges, verification prompts, memory behaviors, notification systems, vendor explanations, and deletion interfaces. When these elements are inconsistent or weakly coordinated, users experience not only confusion but a generalized sense that the system is extracting from them while withholding meaningful control.

This has concrete design and policy implications. First, platforms should reduce \emph{disproportionate entry requirements} by ensuring that identity verification and broad integrations are proportionate, clearly justified, and not silently transformed into conditions for relational participation. Second, they should treat \emph{intensified sensitivity in intimate use} as a signal that later-stage conversations may require different protections than early exploratory use, including clearer options for disabling memory, excluding intimate content from downstream reuse, and selectively forgetting particularly sensitive material, including the learned models, through techniques like model unlearning \cite{bourtoule2021machine}. 
Third, they should address \emph{interpretive uncertainty and perceived surveillance} by making privacy communication coherent across interface cues, policy language, vendor relationships, and actual system behavior rather than relying on fragmented or contradictory signals. Finally, they should treat \emph{irreversibility, persistence, and user burden} as a governance failure mode, supporting clearer distinctions between hiding, archiving, deleting, disabling memory, and preventing future contact, while recognizing that privacy choices are often weak, manipulative, or labor-intensive in practice \cite{schaub2015,bosch2016,gunawan2021,schaffner2022}.

A broader implication is that companion-AI privacy governance should be staged rather than front-loaded. One-time notice and consent at onboarding are poorly matched to systems where the most sensitive disclosures often occur later, after trust and attachment have formed. A lifecycle perspective instead suggests privacy governance that evolves with use: proportionate access requirements at entry, contextual reminders and controls during intimate use, coherent explanations when data practices change, and meaningful reversibility at exit. 

\section{Conclusion}

In this paper, we examined privacy in companion-AI platforms through a qualitative analysis of public Reddit discussions. Analyzing 2,909 posts from 79 subreddits collected over a one year period, we showed that privacy concerns in these systems are better explained as an evolving problem across the lifecycle of platform use, rather than distinct concerns based on specific interactions.
Our findings identified four recurring patterns through which privacy concerns emerge and intensify: disproportionate entry requirements, intensified sensitivity in intimate use, interpretive uncertainty and perceived surveillance, and irreversibility, persistence, and user burden. Together, these themes show that romantic AI users experience privacy as something shaped not only by what they share, but by when platform demands and concerns appear in the duration of use, from installation to termination.
While prior privacy frameworks help explain inappropriate data demands, vulnerable disclosure, and opaque signaling, our findings suggest that deletion failure and exit difficulties reveal a broader lifecycle governance problem that extends beyond instant interaction into retention, reversibility, and user burden.
Our discussion further showed that these issues have implications beyond privacy in a narrow informational sense. Because users often bring emotionally vulnerable, grief-related, therapeutic, or otherwise highly intimate disclosures into companion-AI systems, privacy protections must account for the changing sensitivity of data over time rather than relying only on front-loaded notice and consent. At the same time, failures of deletion, weak reversibility, and inconsistent privacy communication suggest that trust and safety in romantic AI must also include the governance of intimate inputs, memory systems, retention, and exit pathways.



 \begin{acks}

This work is supported by the University of Texas System Rising STARs Award
(No. 40071109), the startup funding from the University
of Texas at Dallas, and a fellowship from the Institute of Health Emergencies and Pandemics at the University of Toronto.
\end{acks}

\bibliographystyle{ACM-Reference-Format}
\bibliography{sample-base}


\end{document}